# Carbon Monitor, a near-real-time daily dataset of global CO2 emission from fossil fuel and cement production

Zhu Liu[1*†], Philippe Ciais[2*†], Zhu Deng[1†], Steven J. Davis[3*], Bo Zheng[2], Yilong Wang[4], Duo Cui[1], Biqing Zhu[1], Xinyu Dou[1], Piyu Ke[1], Taochun Sun[1], Rui Guo[1], Olivier Boucher[5], François-Marie Bréon[2], Chenxi Lu[1], Runtao Guo[6], Eulalie Boucher[7], Frederic Chevallier[2]

[1] Department of Earth System Science, Tsinghua University, Beijing 100084, China.
[2] Laboratoire des Sciences du Climat et de l'Environnement LSCE, Orme de Merisiers 91191 Gif-sur-Yvette, France
[3] Department of Earth System Science, University of California, Irvine, 3232 Croul Hall, Irvine, CA 92697-3100, USA
[4] Key Laboratory of Land Surface Pattern and Simulation, Institute of Geographical Sciences and Natural Resources Research, Chinese Academy of Sciences, Beijing, China
[5] Department of Electrical Engineering, Tsinghua University, Beijing 100084, China
[5] Institute Pierre-Simon Laplace, Sorbonne Université / CNRS, Paris, France
[6] School of Mathematical School, Tsinghua University, Beijing 100084, China
[7] Université Paris-Dauphine, PSL, Paris, France
* Corresponding authors: zhuliu@tsinghua.edu.cn, philippe.ciais@lsce.ipsl.fr, sjdavis@uci.edu
† Authors contribute equally

## Abstract

We constructed a near-real-time daily $CO_2$ emission dataset, namely the Carbon Monitor, to monitor the variations of $CO_2$ emissions from fossil fuel combustion and cement production since January 1st 2019 at national level with near-global coverage on a daily basis, with the potential to be frequently updated. Daily $CO_2$ emissions are estimated from a diverse range of activity data, including: hourly to daily electrical power generation data of 29 countries, monthly production data and production indices of industry processes of 62 countries/regions, daily mobility data and mobility indices of road transportation of 416 cities worldwide. Individual flight location data and monthly data were utilised for aviation and maritime transportation sectors estimates. In addition, monthly fuel consumption data that corrected for daily air temperature of 206 countries were used for estimating the emissions from commercial and residential buildings. This Carbon Monitor dataset manifests the dynamic nature of $CO_2$ emissions through daily, weekly and seasonal variations as influenced by workdays and holidays, as well as the unfolding impacts of the COVID-19 pandemic. The Carbon Monitor near-real-time $CO_2$ emission dataset shows a 7.8% decline of $CO_2$ emission globally from Jan 1st to Apr 30th in 2020 when compared with the same period in 2019, and detects a re-growth of $CO_2$ emissions by late April which are mainly attributed to the recovery of economy activities in China and partial easing of lockdowns in other countries. Further, this daily updated $CO_2$ emission dataset could offer a range of opportunities for related scientific research and policy making.

**Background & Summary**

The main cause of global climate change is the excessive anthropogenic emission of $CO_2$ to the atmosphere from geological carbon reservoirs, from the combustion of fossil fuel and cement production. Dynamic information on fossil fuel-related $CO_2$ emissions is critical for understanding the impacts on climate due to different human activities, and their variability, on the forcing of climate change. Further, the combustion processes of fossil fuel also emit short-lived pollutants such as $SO_2$, $NO_2$ and $CO$. Therefore, such information would also allow a more accurate quantification and better understanding of air quality changes[1,2]. Estimates of $CO_2$ emissions from fossil fuel combustion and cement production[2-8] are based on both activity data (e.g., the amount of fuel burnt or energy produced) and emission factors (See Methods)[9]. The sources of these data are mainly national energy statistics, although a number of organizations such as CDIAC, BP, EDGAR, IEA and GCP also produce and compile estimates for different groups of countries or for all countries[1,10-12]. The reported fossil fuel-related $CO_2$ emissions are usually on an annual basis while lagging the very year's emissions by at least one year.

The uncertainty associated with $CO_2$ emissions from burning fossil fuel and producing cement is small when considering large emitters or the global totals, smaller than that of co-emitted combustion-related pollutants for which uncertain technological factors influence the ratio of emitted amounts to fossil fuel burnt [13-15]. The uncertainty of global carbon emissions from fossil fuel burning and cement production varies between ±6% and ±10%[5,7,16,17] (±2σ). The uncertainty is attributed to both the activity data and the emission factors. For the activity data, the amount of fuel burnt is recorded by energy production and consumption statistics, hence the uncertainties are introduced by errors and inconsistencies in the reported figures from different sources. For the emission factors, the different fuel types, quality and combustion efficiency together contribute to the overall uncertainty. For example, coal used in China is of variable quality and so is its emission factors, both before (raw coal) and after cleaning (cleaned coal) varies significantly, which was found to cause a 15% uncertainty range for $CO_2$ emissions. On the other hand, there is very limited temporal change of emission factors. For example, the annual difference of emission factors for coal consumption was within 2% globally[18] while the variation of emission factors for oil and gas was found to be much smaller.

Given the fact that the uncertainty of $CO_2$ emissions from fossil fuel burning and cement production is in general under ±10% [10,19,20], and the annual difference of emission factors is less than 2% [18], the $CO_2$ emission thus can be estimated directly by estimating the absolute amount and the relative change of activity through time. This method has been widely used for scientific products that update recent changes of $CO_2$ emissions estimates[1,21,22 23], understanding that official and comprehensive $CO_2$ national inventories reported by countries to the UNFCCC become available with a lag of two years for Annex-I countries and several years for non-Annex-I ones[24]. As such, a higher spatial, temporal and sectoral resolution of $CO_2$ emission inventories beyond annual and national level can be obtained by spatial, temporal and sectoral data to disaggregate the annual national emissions[9,14,23,25]. The level of granularity of spatially explicit dynamic emission inventories depends on available data, such as location and operations of point sources[23] (i.e. power generation for a certain plant), regional statistics of energy use (i.e. monthly fuel consumption)[9,25], and knowledge of proxies for the distribution of emissions such as gridded population density, night lights, urban forms and GDP data etc.[9,14,23,25].

Gaining from past experiences of constructing annual inventories and newly compiled activity data, we present in this study a novel daily dataset of $CO_2$ emissions from fossil fuel burning and cement production at national level. The countries/regions include China, India, U.S., Europe (EU27 & UK), Russia, Japan, Brazil, and rest of world (ROW), as well as the emissions from international bunkers. This dataset, known as Carbon Monitor, is separated into several key emission sectors: power sector (39% of total emissions), industrial production (28%), ground transport (18%), air transport (3%), ship transport (2%), and residential consumption (10 %). For the first time, daily emissions estimates are produced for these six sectors, based on dynamically and regularly updated activity data. This is made possible by the availability of recent activity data such as hourly electrical power generation, traffic indices, airplane locations and natural gas distribution, with the assumption that the daily variation of emissions is driven by the activity data and that the contribution from emission factors is negligible, as they evolve at longer time scales, e.g. from policy implementation and technology shifts.

The framework of this study is illustrated in Fig 1. We calculated national $CO_2$ emissions and international aviation and shipping emissions since the Jan 1st 2019, drawing on hourly datasets of electricity power production and their $CO_2$ emissions in 29 countries (thus including the substantial variations in carbon intensity associated with the variable mix of electricity production), daily vehicle traffic indices in 416 cities worldwide, monthly production data for cement, steel and other energy intensive industrial products in 62 countries/regions, daily maritime and aircraft transportation activity data, and either previous-year fuel use data corrected for air temperature to residential and the commercial buildings. Together, these data cover almost all fossil fuels and industry sources of global $CO_2$ emissions, except for the emission from land use change (up to 10% of global $CO_2$ emissions) and non-fossil fuel $CO_2$ emissions of industrial products (up to 2% of global $CO_2$ emissions)[26] in addition to cement and clinker (i.e. plate glass, ammonia, calcium carbide, soda ash, ethylene, ferroalloys, alumina, lead and zinc etc.).

While daily emission can be directly calculated using near-real-time activity data and emission factors for the electricity power sector, such an approach is difficult to apply to all sectors. For the industry sector, emissions can be estimated monthly in some countries. For the other sectors, we used proxy data instead of daily real activity data, to dynamically downscale the annual or monthly $CO_2$ emissions totals on a daily basis. For instance, traffic indices in cities representative of each country were used instead of actual vehicle counts and categories, combined with annual national total sectoral emissions, to produce daily road transportation emissions. As such, for the road transportation, air transportation and residential use of fuels sectors in most countries, we downscaled monthly or annual total emission data in 2019 to calculate the daily $CO_2$ emission in the very year. Subsequently, we scaled monthly totals of 2019 by daily proxies of activities to obtain daily $CO_2$ emissions data in the first four months of 2020, during the unprecedented disturbance of the COVID-19 pandemic. The Carbon Monitor near-real-time $CO_2$ emission dataset shows a 7.8% decline of $CO_2$ emission globally from January 1st to April 30th in 2020 when compared with the same period in 2019, and detects a re-growth of CO2 emissions by late April which are mainly attributed to the recovery of economy activities in China and partial easing of lockdowns in other countries.

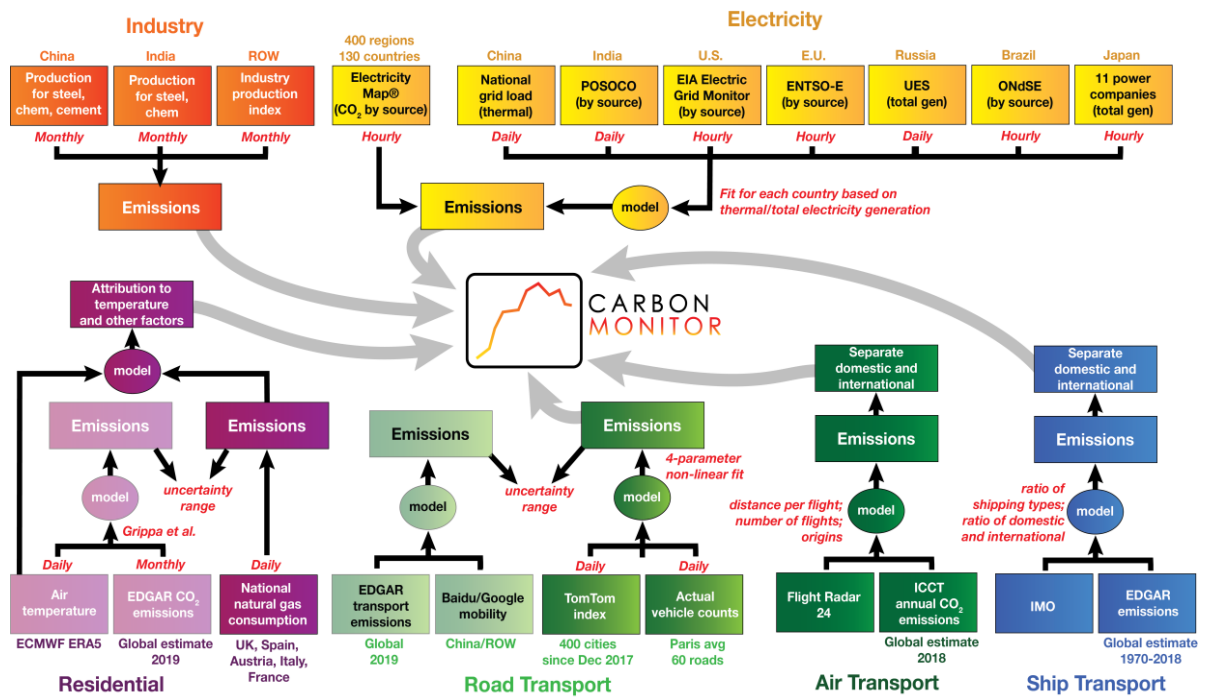

Fig 1. **Framework for data processing**

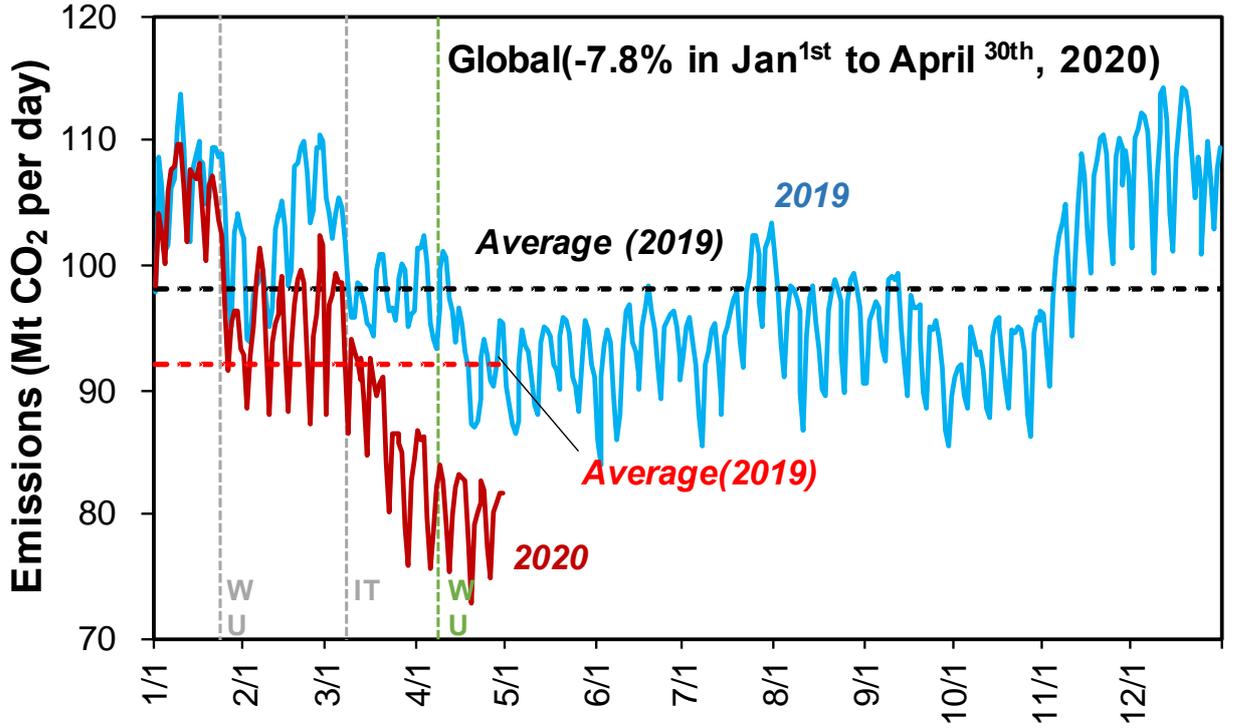

Fig 2. **Daily CO$_2$ emission data from January 1$^{st}$ to April 30$^{th}$ 2020**

## Methods

### Annual total and sectorial emission per country in the baseline year 2019

According to the IPCC Guidelines for emission reporting[4], the CO$_2$ emissions *Emis* should be calculated by multiplying activity data *AD* by corresponding emission factors *EF*.

$$Emis = \sum\sum\sum AD_{i,j,k} \cdot EF_{i,j,k} \qquad (1)$$

Where $i, j, k$ are indices for regions, sectors and fuel types respectively. *EF* can be further separated into the net heating values $v$ for each fuel type (the energy obtained per unit of fuel), the carbon content $c$ per energy output (t C/TJ) and the oxidization rate $o$ (the fraction (in %) of fuel oxidized during combustion):

$$Emis = \sum\sum\sum AD_{i,j,k} \cdot (v_{i,j,k} \cdot c_{i,j,k} \cdot o_{i,j,k}) \qquad (2)$$

Due to the lag of more than two years in publishing governmental energy statistics, we started from the latest CO$_2$ emissions estimates up to 2018 from current CO$_2$ databases[1,10-12]. For 2019, we completed this information to obtain annual total emissions based on literature data and disaggregated the annual total into daily emissions (see below). For 2020, we estimated daily CO$_2$ emissions by using daily changes of activity data in 2020 compared to 2019. The CO$_2$ emissions and sectoral structure in 2018 for countries and regions were extracted from EDGAR V4.3.2[1,27] and V5.0 for each country, and national emissions were scaled to the year

2019 based on our own estimate (for China) and data from the Global Carbon Budget 2019[21] (for other countries):

$$Emis_{r,2019} = \alpha_r \cdot Emis_{r,2018} \qquad (3)$$

For China, we firstly calculated $CO_2$ emissions in 2018 based on the energy consumption by fuel types and cement production in 2018 from China Energy Statistical Yearbook[28] and the National Bureau Statistics[29] following Equation 1. We projected the energy consumption in 2019 from the annual growth rates of coal, oil and gas reported by Statistical Communiqué[29] and applied China-specific emission factors[30] to obtain the annual growth rate of emissions in 2019. For US and Europe (EU27&UK), we used updated emission growth rates in 2019 published by CarbonBrief (https://www.carbonbrief.org/guest-post-why-chinas-co2-emissions-grew-less-than-feared-in-2019). For countries with no estimates of emission growth rates in 2019 such as Russia, Japan and Brazil, we assumed their growth rates of emissions was 0.5% based on the emission growth rate of the rest of world[22].

In this study, the EDGAR sectors were aggregated into four sectors ($s$): power sector, industry sector, transport sector (ground transport, aviation and shipping), and residential sector. This is consistent with the new activity data we used below to compute daily variations. We used the sectoral distribution in 2018 from EDGAR to infer the sectoral emissions in 2019 for each country/region (Equation 4), assuming that the sectoral distribution remained unchanged in these two years.

$$Emis_{r,s,2019} = Emis_{r,2019} \cdot \frac{Emis_{r,s,2018}}{Emis_{r,2018}} \qquad (4)$$

Table 1 **Scaling factor for the emission growth in 2019 compared to 2018**

| Countries/Regions | Scaling Factor (%) | Source |
|---|---|---|
| China | 2.8% | Estimated in this study |
| India | 1.8% | Global Carbon Budget 2019[22] |
| US | 2.4% | Carbon Brief, 2020 |
| EU27&UK | -3.9% | Carbon Brief, 2020 |
| Russia | 0.5% | = ROW |
| Japan | 0.5% | = ROW |
| Brazil | 0.5% | = ROW |
| ROW | 0.5% | Global Carbon Budget 2019[22] |

**Data acquisition and processing of Carbon Monitor daily $CO_2$ emissions**

According to IPCC Guidelines[4], the $CO_2$ emissions for sector could be calculated by multiplying sectoral activity data by their corresponding emission factors following Equation 5:

$$Emis_s = AD_s \cdot EF_s \qquad (5)$$

The emissions were here calculated following this equation, separately for the power sector,

the industry sector, the transport sector, and the residential sector.

**1. Power sector.**

The $CO_2$ emissions from power sector can be calculated by adapting Equation 5 with sector specific activity data (i.e. electricity production/thermal electricity production) and corresponding emission factors (Equation 6):

$$Emis_{power} = AD_{power} \cdot EF_{power} \qquad (6)$$

Normally the emission factors change slightly over time but can be assumed to remain constant over the two years period considered in this study, compared to the huge changes in activity data. Thus, we assumed that emission factors remained unchanged in 2019 and 2020, and calculated the daily emissions as follows:

$$Emis_{daily} = Emis_{yearly} \cdot \frac{AD_{daily}}{AD_{yearly}} \qquad (7)$$

The data sources of daily activity data in power sector are described as Table 2. The countries/regions listed in Table 2 account for more than 70% of the total $CO_2$ emissions in the power sector. For emissions from other countries (ROW), which are not listed in Table 2, we estimated the power sector emission changes in 2020 based on the period of the national lock-down. For daily emission changes of ROW in 2019, we firstly assumed a linear relationship between daily global emission and daily total emissions of the ROW countries listed in Table 2. Then we classified each country according to whether they adopted lock-down measures, based on official reports. Based on daily emission data of the power sector of the countries listed in Table 2, we calculated the respective average change rates of power sectors in ROW countries between January and April, assuming changes started since the date of lock-down in each country. Emissions from countries with no lock-down were left unchanged. We then applied these country-specific January to April emissions growth rates to estimate daily changes for each ROW country in 2020, based on their lock-down measures, and aggregated them into daily emission for ROW.

Table 2 **Data sources of activity data in power sector**

| Country/Region | Data source | Sectors included | Resolution |
| --- | --- | --- | --- |
| China | National Grid Daily Electric Load | Thermal production | Daily |
| India | Power System Operation Corporation Limited (https://posoco.in/reports/daily-reports/) | Thermal production (summarizing the production of *Coal, Lignite*, and *Gas, Naphtha & Diesel*) | Daily |
| US | Energy Information Administration's (EIA) Hourly Electric Grid Monitor (https://www.eia.gov/beta/electricity/grid monitor/) | Thermal production (summarizing the production of *Coal, Petroleum,* and *Natural* | Hourly |
| EU27 & UK | ENTSO-E Transparent platform (https://transparency.entsoe.eu/dashboard/show) | Thermal production (summarizing the production of *Fossil.Brown.coal.Lignite, Fossil.Coal.derived.gas, Fossil.Gas, Fossil.Hard.coal, Fossil.Oil, Fossil.Oil.shale,* and *Fossil.Peat.*) | Croatia, Cyprus, Ireland, Luxembourg and Malta excluded due to unsatisfactory data quality or missing data |
| Russia | United Power System of Russia (http://www.so-ups.ru/index.php) | Total generation | Hourly |
| Japan | Summarizing electricity data from 10 electricity providers in Japan (Hokkaido Electric Power, Tohoku Electric Power Network, Tokyo Electric Power Company, Chubu Electric Power Grid, Hokuriku Electric Power Transmission & Distribution Company, Kansai Electric Power, Chugoku Electric Power Company, Shikoku Electric Power Company, Kyushu Electric Power and Okinawa Electric Power Company). | Total generation | Hourly |
| Brazil | Operator of the National Electricity System (http://www.ons.org.br/Paginas/). | Thermal production | Hourly |

## 2. Industry sector: Industrial and cement production

While daily production data is not directly available for industrial and cement production, the monthly $CO_2$ emissions from industry and cement production sector could be calculated by using monthly statistics of industrial production, and daily data of electricity generation to disaggregate the monthly $CO_2$ emissions into daily values. This calculation assumes a linear relationship between daily electricity generation for industry and daily industry production data to compute daily industry production.

The emissions from industrial production during the fossil fuel combustion were calculated by multiplying activity data (i.e., fossil fuel consumption data in the industrial sector) by corresponding emission factors by type of fuel. Due to limited data availability, we assumed a linear relationship between daily industrial production and industrial fossil fuel use, and the emission factors remaining unchanged. So, the monthly emissions in 2019 in country/region could be calculated by following equation:

$$Emis_{montly,2019,r} = Emis_{yearly,2019,r} \cdot (P_{monthly,2019,r}/P_{yearly,2019,i,r}) \qquad (8)$$

The emissions from cement production during the chemical process of calcination of calcite were calculated with the same Eq.(8), which is normally used by multiplying the cement production by the emission factor of this industry.

Specifically, for China, the emissions from the industry sector were further divided into steel

industry, cement industry, chemical industry, and other industries (indicated by index $i$):

$$Emis_{montly,2019,China} = \sum Emis_{yearly,2019,i} \cdot (P_{monthly,2019,i}/P_{yearly,2019,i}) \quad (9)$$

For the monthly emissions in 2020 in country/region , we used the following equation:

$$Emis_{montly,2020,r} = Emis_{monthly,2019,r} \cdot (P_{monthly,2020,r}/P_{monthly,2019,r}) \quad (10)$$

where $P$ is the industrial production in different industrial sectors (in China) or a total Industrial Production Index (in other countries) as listed in Table 3. In China's case, the January and February estimates were combined as no individual monthly data was reported by sources listed in Table 3 for these two months. The monthly industrial emissions were disaggregated to daily emissions using daily electricity data, as explained above.

Lacking in the latest Industrial Production Index in April 2020 for Europe, India, Japan, Russia and Brazil, we adopted monthly growth rates of industrial output from Trading Economics (https://tradingeconomics.com) based on preliminary survey data. For other countries not listed in Table 3, we used the same method as described for the power sector to calculate the daily industry emissions from ROW.

To allocate monthly emissions into daily emissions, we use the weight of daily electricity production to monthly electricity production:

$$Emis_{daily} = Emis_{monthly} \cdot (Elec_{daily}/Elec_{monthly}) \quad (11)$$

Table 3 **Data sources for industrial production**

| Country/Region | Sector | Data | Data source |
|---|---|---|---|
| China | Steel industry | Crude steel production | World Steel Association website (https://www.worldsteel.org/) |
| | Cement Industry | Cement and clinker production | National Bureau of Statistics (http://www.stats.gov.cn/english/) |
| | Chemical industry | sulfuric acid, caustic soda, soda ash, ethylene, chemical fertilizer, chemical pesticide, primary plastic and synthetic rubber | National Bureau of Statistics (http://www.stats.gov.cn/english/) |
| | Other industry | crude iron ore, phosphate ore, salt, feed, refined edible vegetable oil, fresh and frozen meat, milk products, liquor, soft drinks, wine, beer, tobaccos, yarn, cloth, silk and woven fabric, machine-made paper and paperboards, plain glass, ten kinds of nonferrous metals, refined copper, lead, zinc, electrolyzed aluminum, industrial boilers, metal smelting equipment, and cement equipment | National Bureau of Statistics (http://www.stats.gov.cn/english/) |
| India | / | Industrial Production Index (IPI) | Ministry of Statistics and Programme Implementation |
| | | | (http://www.mospi.nic.in) |
| US | / | Industrial Production Index (IPI) | Federal Reserve Board (https://www.federalreserve.gov) |
| EU & UK | / | Industrial Production Index (IPI) | Eurostat (https://ec.europa.eu/eurostat/home) |
| Russia | / | Industrial Production Index (IPI) | Federal State Statistics Service (https://eng.gks.ru) Trading Economics (https://tradingeconomics.com) |
| Japan | / | Industrial Production Index (IPI) | Ministry of Economy, Trade and Industry (https://www.meti.go.jp) Trading Economics (https://tradingeconomics.com) |
| Brazil | / | Industrial Production Index (IPI) | Brazilian Institute of Geography and Statistics (https://www.ibge.gov.br/en/institutional/the-ibge.htm) Trading Economics (https://tradingeconomics.com) |

### 3. Transport sector

#### 1) Road transportation

We collected hourly TomTom congestion level data from the TomTom website (https://www.tomtom.com/en_gb/traffic-index/). The congestion level (called $X$ hereafter) represents the extra time spent on a trip, in percentage, compared to uncongested condition. TomTom congestion level data were obtained for 416 cities across 57 countries at a temporal resolution of one hour. Of note that a zero-congestion level means that the traffic is fluid or 'normal', but does not mean there was no vehicle and zero emissions. It is thus important to identify the lower threshold of emissions when the congestion level is zero. To do so, we compared the time series of daily mean TomTom congestion level $X$, with the daily mean car flux (called hereafter in vehicle per day) from publicly available real-time $Q$ data from an average of 60 roads in the Paris megacity. Those daily mean car counts were reported by the City's service (https://opendata.paris.fr/pages/home/). We used a sigmoid function to fit the relationship between $X$ and $Q$ (Fig 2):

$$Q = a + \frac{bX^c}{d^c + X^c} \quad (12)$$

where a, b, c and d are the regression parameters (Table 4). We verified that the empirical fit from Eq. (12) can reproduce the observed large drop of $Q$ due to the lockdown in Paris and the recovery afterwards. We assume that daily emissions relative changes were proportional to the relative change of the function $Q(X)$ from Eq. (12). Then, we applied the function $Q(X)$ established for Paris to other cities included in the TomTom dataset, assuming that the relative magnitude in car counts (and thus emissions) follow similar relationship with TomTom. The emission changes were first calculated for individual cities, and then weighted by city emissions to aggregate to national changes. For a specific country $i$ with $n$ cities reported by TomTom, the national daily vehicle flux for day $j$ was given by:

$$Q_{country,dayj} = \frac{\sum_{i=1}^{n} Q_{i,dayj} E_i}{\sum_{i=1}^{n} E_i} \qquad (13)$$

Where is the annual road transportation emission of city $n$ taken in the grid point of each TomTom city from the annual gridded EDGARv4.3.2 emission map for the "road transportation" sector (1A3b) (httpiis://edgar.jrc.ec.europa.eu/) for the year 2010, assuming that the spatial distribution of ground transport did not change significantly within a country between 2010 and the period of this study. Then, the daily road transportation emissions in 2019 and 2020 ($E_{country,dayj}$) for a country were scaled such that the total road transportation emissions in the first four months of 2019 equaled to 121/365 times the annual emissions of this sector in 2019 ($E_{country,2019}$) estimated in this study:

$$E_{country,dayj} = Q_{country,dayj} \frac{121/365 \times E_{country,2019}}{\sum_{j=1}^{121} Q_{country,dayj}(2019)} \qquad (14)$$

For countries not included in the TomTom dataset, we assumed that the emission changes follow the mean changes of other countries. For example, Cyprus, as an EU member country, had no city reported in TomTom dataset, so its relative emission change was assumed to follow the same pattern of the total emissions from other EU countries included in TomTom dataset (which covers 98% of EU total emissions). Similarly, the relative emission changes of countries in ROW but not reported by TomTom were assumed to follow the same pattern of the total emissions from all TomTom reported countries (which cover 85% of global total emissions).

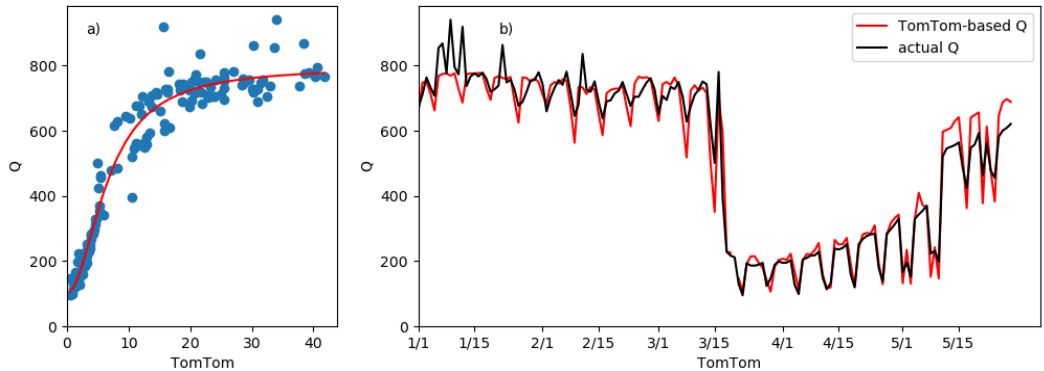

Fig 3 **a) The relationship between TomTom congestion level index ($X$) and actual car counts ($Q$) for Paris. The sigmoid fit between $X$ and $Q$ is given by the red line. b) evaluation of the function $Q(X)$ during the period of the lock down in Paris.**

Table 4 **Regression parameters of the sigmoid function of Eq. (12) that describes the relationship between car counts ($Q$) and TomTom congestion level ($X$)**

| Parameter | Value |
|---|---|
| $a$ | 100.87 |
| $b$ | 671.06 |
| $c$ | 1.98 |
| $d$ | 6.49 |

Table 5 **Cities (416 across 57 countries) where TomTom congestion level data is available**

| Country/Region | City |
|---|---|
| Austria (5) | Vienna, Salzburg, Graz, Innsbruck, Linz |
| Belgium (10) | Brussels, Antwerp, Namur, Leuven, Ghent, Liege, Kortrijk, Mons, Bruges, Charleroi |
| Bulgaria (1) | Sofia |
| Czech (3) | Brno, Prague, Ostrava |
| Denmark (3) | Copenhagen, Aarhus, Odense |
| Estonia (1) | Tallinn |
| Finland (3) | Helsinki, Turku, Tampere |
| France (25) | Paris, Marseille, Bordeaux, Nice, Grenoble, Lyon, Toulon, Toulouse, Montpellier, Nantes, Strasbourg, Lille, Clermont-Ferrand, Brest, Rennes, Rouen, Le-havre, Saint-Etienne, Nancy, Avignon, Orleans, Le-mans, Dijon, Reims, Tours |
| Germany (26) | Hamburg, Berlin, Nuremberg, Bremen, Stuttgart, Munich, Bonn, Frankfurt-am-main, Dresden, Cologne, Wiesbaden, Ruhr-region-west, Leipzig, Hannover, Kiel, Freiburg, Dusseldorf, Karlsruhe, Ruhr-region-east, Munster, Augsburg, Monchengladbach, Mannheim, Bielefeld, Wuppertal, Kassel |
| Greece (2) | Athens, Thessaloniki |
| Hungary (1) | Budapest |
| Iceland (1) | Reykjavik |
| Ireland (3) | Dublin, Cork, Limerick |
| Italy (25) | Rome, Palermo, Messina, Genoa, Naples, Milan, Catania, Bari, Reggio-calabria, Bologna, Florence, Turin, Prato, Cagliari, Pescara, Livorno, Trieste, Verona, Taranto, Reggio-emilia, Ravenna, Padua, Parma, Modena, Brescia |
| Latvia (1) | Riga |
| Lithuania (1) | Vilnius |
| Luxembourg (1) | Luxembourg |
| Netherlands (17) | The-hague, Haarlem, Leiden, Arnhem, Amsterdam, Rotterdam, Nijmegen, Groningen, Eindhoven, Utrecht, Amersfoort, Tilburg, Breda, Apeldoorn, Zwolle, Den-bosch, Almere |
| Norway (4) | Oslo, Trondheim, Stavanger, Bergen |
| Poland (12) | Lodz, Krakow, Poznan, Warsaw, Wroclaw, Bydgoszcz, Gdansk-gdynia-sopot, Szczecin, Lublin, Bialystok, Bielsko-biala, Katowice-urban-area |
| Portugal (5) | Lisbon, Porto, Funchal, Braga, Coimbra |
| Romania (1) | Bucharest |
| Russia (11) | Moscow, Saint-petersburg, Novosibirsk, Yekaterinburg, Nizhny-novgorod, Samara, Rostov-on-don, Chelyabinsk, Omsk, Tomsk, Kazan |
| Slovakia (2) | Bratislava, Kosice |
| Slovenia (1) | Ljubljana |
| Spain (25) | Barcelona, Palma-de-mallorca, Granada, Madrid, Santa-cruz-de-tenerife, Seville, A-coruna, Valencia, Malaga, Murcia, Las-palmas, Alicante, Santander, Pamplona, Gijon, Cordoba, Zaragoza, Vitoria-gasteiz, Vigo, Cartagena, Valladolid, Bilbao, Oviedo, San-sebastian, Cadiz |
| Sweden (4) | Stockholm, Uppsala, Gothenburg, Malmo |
| Switzerland (6) | Geneva, Zurich, Lugano, Lausanne, Basel, Bern |

| Country | Cities |
|---|---|
| Turkey (10) | Istanbul, Ankara, Izmir, Antalya, Bursa, Adana, Mersin, Gaziantep, Konya, Kayseri |
| Ukraine (4) | Kiev, Odessa, Kharkiv, Dnipro |
| UK (25) | Edinburgh, London, Bournemouth, Hull, Belfast, Brighton-and-hove, Bristol, Manchester, Leicester, Coventry, Nottingham, Cardiff, Birmingham, Southampton, Leeds-bradford, Liverpool, Sheffield, Swansea, Newcastle-sunderland, Glasgow, Reading, Portsmouth, Stoke-on-trent, Preston, Middlesbrough |
| Egypt (1) | Cairo |
| South Africa (6) | Cape-town, Johannesburg, Pretoria, East-london, Durban, Bloemfontein |
| China (22) | Chongqing, Zhuhai, Guangzhou, Beijing, Chengdu, Changchun, Changsha, Shenzhen, Shenyang, Shanghai, Wuhan, Fuzhou, Shijiazhuang, Xiamen, Nanjing, Hangzhou, Tianjin, Ningbo, Quanzhou, Dongguan, Suzhou, Wuxi |
| Hong Kong (1) | Hong Kong |
| India (4) | Mumbai, New-delhi, Bangalore, Pune |
| Indonesia (1) | Jakarta |
| Israel (1) | Tel-aviv |
| Japan (5) | Tokyo, Osaka, Nagoya, Sapporo, Kobe |
| Kuwait (1) | Kuwait-city |
| Malaysia (1) | Kuala-lumpur |
| Philippines (1) | Manila |
| Saudi Arabia (2) | Riyadh, Jeddah |
| Singapore (1) | Singapore |
| Taiwan (5) | Kaohsiung, Taipei, Taichung, Tainan, Taoyuan |
| Thailand (1) | Bangkok |
| United Arab Emirates (2) | Dubai, Abu-dhabi |
| Australia (10) | Sydney, Melbourne, Brisbane, Adelaide, Gold-coast, Hobart, Newcastle, Perth, Canberra, Wollongong |
| New Zealand (6) | Auckland, Wellington, Hamilton, Christchurch, Dunedin, Tauranga |
| Argentina (1) | Buenos-aires |
| Brazil (9) | Recife, Sao-paulo, Rio-de-janeiro, Salvador, Fortaleza, Porto-alegre, Belo-horizonte, Curitiba, Brasilia |
| Chile (1) | Santiago |
| Columbia (1) | Bogota |
| Peru (1) | Lima |
| Canada (12) | Vancouver, Toronto, Montreal, Ottawa, London, Winnipeg, Halifax, Quebec, Hamilton, Calgary, Edmonton, Kitchener-waterloo |
| Mexico (1) | Mexico-city |
| USA (80) | Los-angeles, New-york, San-francisco, San-jose, Seattle, Miami, Chicago, Washington, Honolulu, Atlanta, Baton-rouge, San-diego, Boston, Austin, Portland, Philadelphia, Sacramento, Houston, Riverside, Tampa, Nashville, Orlando, Charleston, Denver, Cape-coral-fort-myers, Pittsburgh, New-orleans, Las-vegas, Boise, Fresno, Baltimore, Tucson, Providence, Charlotte, Dallas-fort-worth, Oxnard-thousand-oaks-ventura, Bakersfield, Greenville, Jacksonville, Detroit, Albuquerque, Columbus, San-antonio, Salt-lake-city, Phoenix, Mcallen, Raleigh, Virginia-beach, Hartford, Colorado-springs, Birmingham, New-haven, Louisville, Minneapolis, Cincinnati, El-paso, Allentown, Buffalo, Memphis, Worcester, Grand-rapids, Albany, St-louis, Milwaukee, Omaha-council-bluffs, Indianapolis, Rochester, Columbia, Oklahoma-city, Cleveland, Tulsa, Kansas-city, Knoxville, Richmond, Winston-salem, Dayton, Little-rock, Syracuse, Akron, Greensboro-high-point |

2) **Aviation**

We calculated $CO_2$ emissions from commercial aviation following a commonly used approach: reconstructing the emission inventories from bottom up based on the knowledge of the parameters of individual flights. We collected the FlightRadar24 data (https://www.flightradar24.com/) for the departure and landing airports for each flight, the calculate the distance flown assuming the shortest distance for each flight, and then $CO_2$ emissions per flight[31]. Flights were grouped per country, and for each country between domestic or international traffic. The daily $CO_2$ emission was computed as the product of distance flown, by a $CO_2$ emission factor per *km* flown, according to:

$$Daily\ Emis_{aviation} = Daily\ Kilometers\ Flown_{aviation\ 2020} \times EF_{aviation\ 2019} \quad (14)$$

We acquired monthly individual commercial flight information from FlightRadar24. Individual commercial flights are tracked by FlightRadar24 based on reception of ADS-B signals emitted by aircraft and received by their network of ADS-B receptors[31]. The *Daily Kilometers Flown* are computed assuming great circle distance between the take-off, cruising, descent and landing points for each flight and are cumulated over all flights. As there is no sufficient data available to convert the FlightRadar24 database into $CO_2$ emissions on a flight-by-flight basis, we computed $CO_2$ emissions by assuming a constant $CO_2$ emission factor per km flown across the whole fleet of aircraft (regional, narrowbody passenger, widebody passenger and freight operations). This assumption is justified if the mix of flights between these categories has not changed substantially between 2019 and 2020.

$$\begin{aligned}EF_{aviation\ 2019} = &\ Annual\ Emis_{aviation\ 2018} \\ &\times Growth\ Rate_{aviation\ 2018-2019} \\ &/ Total\ Etimated\ Number\ of\ Kilometers\ Flown_{aviation\ 2019}\end{aligned} \quad (15)$$

EDGAR published an estimate of total $CO_2$ emissions from commercial aviation in 2018 of 925 Mt $CO_2$. And the International Council on Clean Transportation (ICCT) implied annual compound growth rate of total emissions from commercial flights, 5.7%, during the past five years from 2013 to 2018[32]. In the absence of further information, we considered this increase to be representative of the emission growth rate of commercial aviation from 2018 to 2019. The FlightRadar24 database has incomplete data for some flights and may miss altogether a small fraction of actual flights[31], so we scaled the EDGAR estimate of $CO_2$ emissions (inflated by 5.7% for the year 2019) with the total estimated number of kilometers flown in 2019 (67.91 million km) and apply this scaling factor to 2020 data. We assumed that the fraction of missed flights was the same in 2019 and 2020, which is reasonable.

3) **Ships**

We collected international CO2 shipping emissions from 2016-2018 based on the EDGAR's international emissions. We also. collected global shipping emissions during the period of 2007-2015 from IMO[33] and ICCT (https://theicct.org/sites/default/files/publications/Global-

shipping-GHG-emissions-2013-2015_ICCT-Report_17102017_vF.pdf). According to the Third IMO GHG Study[33], CO2 emissions from international shipping accounted for 88% of global shipping emissions, domestic and fishing accounts for 8% and 4%, respectively. We calculated international CO2 shipping emissions from 2007-2015 from global shipping emissions and the ratio of international shipping and global shipping emissions. We extrapolated emissions from linear fits 2007-2018 to estimate the emissions in 2019. The data sources of shipping emissions are in Table 6. We obtained emissions for the first quarter of 2019 based on the assumption the equal distribution of monthly shipping $CO_2$ emissions. The equations are as follows:

$$Monthly\ Emis_{international\ shipping, 2019} = \alpha \times Yearly\ Emis_{international\ shipping, 2019} \times R_{month} \quad (16)$$

$\alpha$ is the increasing rate of international shipping emissions in 2019 based on the linear extrapolation of data from the period 2007-2018, estimated to be of 3.01%. $R_{month}$ represents the ratio of the months to be calculated in the whole year. Given this, we estimated the shipping emissions for the first quarter of 2019, $R_{month}$ equals 121/365.

We assumed that the change in shipping emissions was linearly related to the change in ships. Traffic volume. The change of international shipping emissions for the first four months of 2020 was calculated according to the following equation:

$$Emis_{period, 2020} = Emis_{period, 2019} \times C_{index} \quad (17)$$

Where represents the ratio of the change in shipping emissions, estimated to the end of Apr by -15% compared to the same period of last year according to https://www.theedgemarkets.com/article/global-container-shipments-set-fall-30-next-few-months.

Table 6. **Data sources used to estimate ship emissions**

| Shipping Emissions | Sources |
|---|---|
| Global shipping Emissions (2007-2012) | IMO[33] |
| Global shipping Emissions (2013-2015) | ICCT |
| International shipping Emissions (2016-2018) | EDGAR v5.0 |

### 4. Residential sector: residential and commercial buildings

Fuel consumption daily data from this sector are not available. Several studies (ref) showed that the main source of daily and monthly variability of this sector is climate, namely heating emissions increase when temperature falls below a threshold which depends on region, building types and people habits. We calculated emissions by assuming annual totals unchanged from 2019 and using climate daily climate information, in three steps: 1) estimation of population-weighted heating degree days for each country and for each day based on the ERA5[34] reanalysis of 2-meters air temperature, 2) split residential emissions into two parts: cooking emissions and heating emissions according to the EDGAR database[35], using the EDGAR estimates of 2018 residential emissions as the baseline. Emissions from cooking were assumed to remain independent of temperature, and those from heating were assumed to be a function of the heating demand. Based on the change of population-weighted heating degree days in each country in 2019 and 2020, we downscaled annual EDGAR 2018 residential emissions to daily values for 2019 and 2020 as described by Eq. 18-20:

$$Emis_{c,m} = Emis_{c,m,2018} \times \frac{\sum_m HDD_{c,d}}{\sum_{m,2018} HDD_{c,d}} \quad (18)$$

$$Emis_{c,d} = Emis_{c,m} \times Ratio_{heating,c,m} \times \frac{HDD_{c,d}}{\sum_m HDD_{c,d}} + Emis_{c,m} \times (1 - Ratio_{heating,c,m}) \times \frac{1}{N_m} \quad (19)$$

$$HDD_{c,d} = \frac{\sum(Pop_{grid} \times (T_{grid,c,d} - 18))}{\sum(Pop_{grid})} \quad (20)$$

where $c$ is country, $d$ is day, $m$ is month, $Emis_{c,m}$ is the residential emissions of country $c$ in month $m$ of the year 2019 or 2020, $Emis_{c,m,2018}$ is the emissions of country $c$ in month $m$ of the year 2018, $HDD_{c,d}$ is the population-weighted heating degree day in country $c$ in day $d$, $Emis_{c,d}$ is the residential emissions of country $c$ in day $d$ of the year 2019 or 2020, $Ratio_{heating,c,m}$ is the percentage of residential emissions from heating demand in country $c$ in month $m$, $N_m$ is the number of days in month $m$, $Pop_{grid}$ is gridded population data derived from Gridded Population of the World, Version 4[36], $T$ is the daily average air temperature at 2 meter derived from ERA5[34].

The main assumption is this approach is that residential emissions did not change from other factors than heating degree days variations in 2020, when people time in houses dramatically increased during the lock-down period. In order to test the validity of this assumption, we compiled natural gas daily consumption data by residential and commercial buildings for France (https://www.smart.grtgaz.com/fr/consommation) (unfortunately such data could not be collected in many countries) during 2019 and 2020. Natural gas consumption in kWh per day was transformed to $CO_2$ emissions using an emission factor of 10.55 kWh per $m^3$ and a molar volume of 22.4 $10^{-3}$ $m^3$ per mole.

Firstly, we verified that the temporal variation of those 'true' residential $CO_2$ emissions was similar to that given by equations (18) to (20). Secondly, after fitting a piecewise model to those natural gas residential emission data using ERA5 air temperature data, we removed the effect of temperature to obtain an emission corrected for temperature effects. Even if the lock down was very strict in France, we found no significant emission anomaly, meaning that the fact that nearly the entire population was confined at home did not increase or decrease emissions. This complementary analysis tentatively suggests that residential emissions can be well approximated in other countries by equations (18) to (20) based only on temperature during the lock down period.

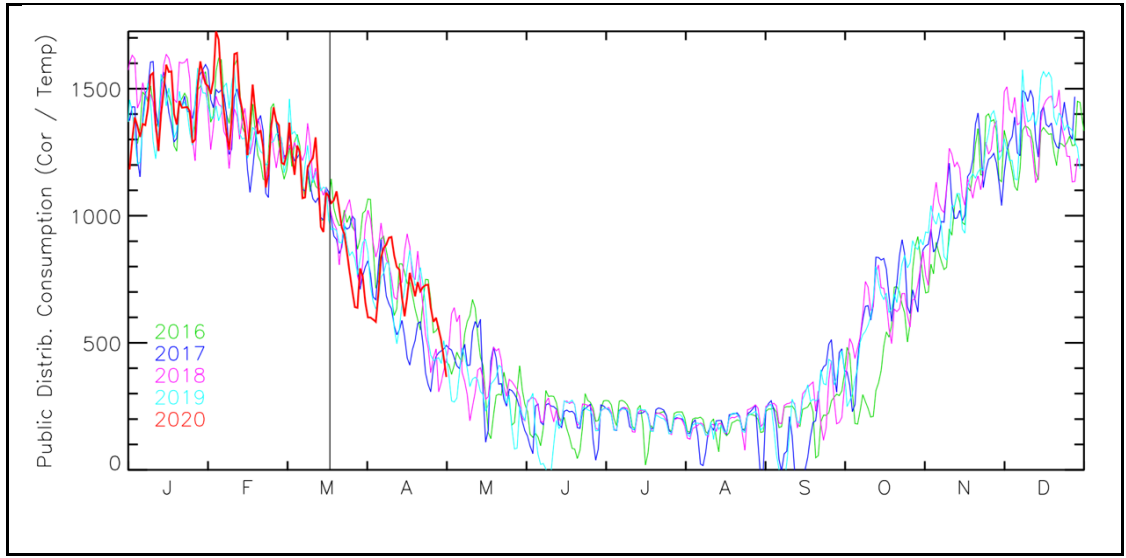

Fig 4. **Residential and commercial building daily natural gas consumption (linearly related to $CO_2$ emissions from this sector) in France for the last 5 years. Temperature effects have been removed from emissions using a linear piecewise model. When the effect of variable winter temperature was accounted for, no significant change is seen in 2020 during the very strict lock-down period.**

**Data Records**

Currently there are 27484 data records provided in this dataset:

- 268 records are daily mean $CO_2$ emissions (from fossil fuel combustion and cement production process) 1751-2020.

- 4374 records are the daily emissions for 9 countries/regions (China, India, US, EU27&UK, Russia, Japan, Brazil, ROW and Globe) and 486 days (from January 1st 2019 to April 30th 2020).

- 22842 records are daily emissions in power sector, ground transport sector, industry sector, residential sector, aviation sector and international shipping sector respectively, for 9 countries/regions (China, India, US, EU27&UK, Russia, Japan, Brazil, ROW and Globe) and 486 days (from January 1st 2019 to April 30th 2020).

**Technical Validation**

**Uncertainty estimates**

We followed the 2006 IPCC Guidelines for National Greenhouse Gas Inventories to conduct the uncertainty analysis of the data. Firstly, the uncertainties were calculated for each sector:

- Power sector: the uncertainty is mainly from inter-annual variability of coal emission factors. Based the UN statistics the inter-annual variability of fossil fuel is within (±1.5%), which been used as uncertainty of the $CO_2$ from power sectors.

- Industrial sector: Uncertainty of $CO_2$ from Industry and cement production comes from the monthly production data. Given $CO_2$ from Industry and cement production in China accounts for more than 60% of world total industrial $CO_2$, and the fact that uncertainty of emission in China is t Uncertainty from monthly statistics was derived from 10000 Monte Carlo simulations to estimate a 68% confidence interval (1-sigma) for China. from monthly statistics was derived from 10000 Monte Carlo simulations to estimate a 68% confidence interval (1-sigma) for China. We calculated the 68% prediction interval of linear regression models between emissions estimated from monthly statistics and official emissions obtained from annual statistics at the end of each year, to deduce the one-sigma uncertainty involved when using monthly data to represent the whole year's change. The squared correlation coefficients are within the range of 0.88 (e.g., coal production) and 0.98 (e.g., energy import and export data), which represent that only using the monthly data can explain 88% to 98% of the whole year's variation[37], while the remaining variation not covered yet reflect the uncertainty caused by the frequent revisions of China's statistical data after they are first published.

- Ground Transportation: The emissions in ground transportation sector is estimated by assuming that the relative magnitude in car counts (and thus emissions) follow the similar relationship with TomTom. So the emissions were quantified by the prediction interval of the regression.

    - Aviation: Uncertainties in the aviation CO2 emissions are difficult to assess. Sources of uncertainties arise from the ICCT (2018) estimate used to scale emissions, the lack of completeness of the flight database and the fixed average conversion factor between kilometre flown and CO2 emissions. These last two uncertainties should have a limited impact as we do not expect a change between 2019 and 2020 in database completeness and in the average fleet composition. In the study the uncertainty of aviation sector comes from the difference of daily emission data estimated based on the two methods. We calculate the average difference between the daily emission results estimated based on the flight route distance and the number of flights, and then divide the average difference by the average of the daily emissions estimated by the two methods to obtain the uncertainty of $CO_2$ from aviation sector.

- Shipping: We used the uncertainty analysis from IMO as our uncertainty estimate for shipping emissions. According to Third IMO Greenhouse Gas study 2014[33], the uncertainty of shipping emissions was 13% based on bottom-up estimates.

- Residential: The 2-sigma uncertainty in daily emissions are estimated as 40%, which is calculated based on the comparison with daily residential emissions derived from real fuel consumptions in several European countries including France, Great Britain, Italy,

Belgium, and Spain.

The uncertainty of emission projection in 2019 is estimated as 2.2%, by combining the reported uncertainty of the projected growth rates and the EDGAR estimates in 2018.

Then we combine all the uncertainties by following the error propagation equation from IPCC. Eq. (21) is used to derive for the uncertainty of the sum, which could be used to combine the uncertainties of all sectors:

$$U_{total} = \frac{\sqrt{\sum(U_s \cdot \mu_s)}}{|\sum \mu_s|} \tag{21}$$

Where $U_s$ and $\mu_s$ are the percentage uncertainties and the uncertain quantities (daily mean emissions) of sector $s$ respectively.

Eq. (22) is used to derive for the uncertainty of the multiplication, which is used to combine the uncertainties of all sectors and of the projected emissions in 2019:

$$U_{overall} = \sqrt{\sum U_i^2} \tag{22}$$

Table 6 **Percentage uncertainties of all items.**

| Items | Uncertainty Range |
|---|---|
| Power | ±1.5% |
| Ground Transport | ±9.3% |
| Industry | ±36.0% |
| Residential | ±40.0% |
| Aviation | ±10.2% |
| International Shipping | ±13.0% |
| Projection of emission growth rate in 2019 | ±0.8% |
| EDGAR emissions in 2018 | ±5.0% |
| **Overall** | **±6.8%** |

**Data Availability Statement**

All data generated or analyzed during this study are included in this article.

**Code Availability**

The code generated during and/or analyzed during the current study are available from the corresponding author. After peer-reviewed the code will be open accessible on the Carbon Monitor website (www.carbonmonitor.org or www.carbonmonitor.org.cn).

**Completing Interests statement**

Authors declare no conflict of interests.

**Author contribution:**

Zhu Liu and Philippe Ciais designed the research, Zhu Deng coordinated the data processing. Zhu Liu, Philippe Ciais and Zhu Deng contributed equally in this research, all authors contributed to data collection, analysis and paper writing.